\documentclass[runningheads]{llncs}
\usepackage{float,amssymb}
\usepackage{graphicx}
\usepackage{subfigure}
\usepackage[colorlinks,linkcolor=blue]{hyperref}
\usepackage{makeidx} 
\usepackage{pifont}

\newsavebox\CBox
\def\textBF#1{\sbox\CBox{#1}\resizebox{\wd\CBox}{\ht\CBox}{\textbf{#1}}}

\begin{document}
\title{EllipseNet: Anchor-Free Ellipse Detection for Automatic Cardiac Biometrics in Fetal Echocardiography}
\titlerunning{EllipseNet}
%
\author{Jiancong Chen\inst{1}\thanks{Authors contributed equally. }  \and Yingying Zhang\inst{2,3\star} \and Jingyi Wang\inst{4,5} \and Xiaoxue Zhou\inst{4,5}  \and \\ Yihua He\inst{4,5\star\star} \and Tong Zhang\inst{1}\thanks{Corresponding authors:  \email{zhangt02@pcl.ac.cn},\email{ heyihuaecho@hotmail.com}. }
}

%
\authorrunning{J. Chen et al.}

\institute{$^1$ Peng Cheng Laboratory, Shenzhen, China; \\ 
    $^2$ School of Biological Science and Medical Engineering, Beihang University, China; \\
    $^3$ Beijing Advanced Innovation Center for Big Data-Based Precision Medicine, China;\\
    $^4$ Echocardiography Medical Center, Beijing Anzhen Hospital, Capital Medical University, China; \qquad
    $^5$ Maternal-Fetal Medicine center in Fetal Heart Disease, Beijing Anzhen Hospital, China
}
%
\maketitle              
\begin{abstract}
As an important scan plane, four chamber view is routinely performed in both second trimester perinatal screening and fetal echocardiographic examinations. The biometrics in this plane including cardio-thoracic ratio (CTR) and cardiac axis are usually measured by sonographers for diagnosing congenital heart disease. However, due to the commonly existing artifacts like acoustic shadowing, the traditional manual measurements not only suffer from the low efficiency, but also with the inconsistent results depending on the operators' skills. In this paper, we present an anchor-free ellipse detection network, namely EllipseNet, which detects the cardiac and thoracic regions in ellipse and automatically calculates the CTR and cardiac axis for fetal cardiac biometrics in 4-chamber view. In particular, we formulate the network that detects the center of each object as points and regresses the ellipses' parameters simultaneously. We define an intersection-over-union loss to further regulate the regression procedure. We evaluate EllipseNet on clinical echocardiogram dataset with more than 2000 subjects. Experimental results show that the proposed framework outperforms several state-of-the-art methods. Source code will be available at \url{https://git.openi.org.cn/capepoint/EllipseNet}.

\end{abstract}
\section{Introduction}
Fetal congenital heart disease (CHD) is one of the most common forms of birth defects worldwide, which affects eight out of 1000 newborn babies worldwide every year~\cite{VanDerLinde2011}. Despite the tremendous efforts made to improve the obstetric screening, the detection of CHD remains a relatively low yield level at 30\% $\sim$ 60\% ~\cite{Bakker2019,Sharland2012}. Four chamber view is a standardized cardiac plane that is regularly scanned in mid-trimester antenatal screening. In 4-chamber view examinations, cardiothoracic diameter ratio (CTR) and cardiac axis are two important and commonly used metrics. The CTR estimates the heart size by measuring the maximum diameter of the cardiac and comparing it to the maximum diameter of the thorax. The cardiac axis is estimated by measuring the angle between the line bisecting the chest and the line along the cardiac interventricular septum. These cardiac biometrics are reliable perimeters in the prediction of multiple CHD~\cite{Sharland2012}. Due to the commonly existing acoustic shadowing artifact in fetal echocardiogram, the traditional manual measurements are very time-consuming and suffer from large inter-operator variance depending on the sonographer's skills. Therefore, automatic and reliable fetal cardiac measurements are in high demand.

With the rapid development of deep learning, convolutional neural networks (CNN) have been applied in many medical related applications and achieved great success for a number of image analysis tasks. Baumgartner et al.~\cite{Baumgartner2017} developed a CNN, named SonoNet, to detect and localize standardised scan planes in freehand ultrasound. Gong et al.~\cite{Gong2020} presented a one-class classification network to distinguish CHD and healthy subjects, where the improved generative adversarial networks were used for data augmentation. Sinclair et al.~\cite{Sinclair2018} developed a segmentation and ellipse fit network for automatic measurement of fetal head circumference and biparietal diameter. Compared to fetal head, fetal echocardiographic measurement is challenged by the moving heart and shadowing artifacts around fetal sternum. 

In this work, we present a one-stage ellipse detection framework for automatic fetal cardiac biometrics in 4-chamber view scans. Unlike~\cite{Sinclair2018}, our method directly detects the objects without the segmentation procedure. Here, we use ellipses to represent the cardiac and thoracic regions, and calculate the CTR and cardiac axis accordingly. Following the popular anchor-free detection networks~\cite{Zhou2019c,Yang2020a}, the proposed EllipseNet detects cardiac and thoracic regions and regresses the ellipses's paramenters in one-stage. The main contributions of our work can be summarized as follows: 

1. The proposed EllipseNet can detect cardiac and thoracic regions and calculate fetal CTR and cardiac axis in 4-chamber view. To the best of our knowledge, this is the first approach to perform cardiac biometrics automatically in fetal echocardiography.

2. EllipseNet is a one-stage ellipse detection framework. Compared to the traditional bounding box based detection, modeling the object with an ellipse better depicts a number of medical structures.

3. A rotated intersection over union (IoU) loss is introduced in ellipse regression, which
significantly improves the network performance in precision of object localization.

\subsection{Related Works}

\textBF{Ellipse detection.} Due to its superior presentation for objects, ellipses have been investigated in the object detection communities. Li et al.~\cite{Li2019c} replaced the Region Proposal Network (RPN) in Faster R-CNN~\cite{DBLP:conf/nips/RenHGS15} by a Gaussian proposal network (GPN), which generates elliptical proposals and represents these ellipses by 2D Gaussian distributions. Pan et al.~\cite{Pan_2021_WACV} further modified the loss with Wasserstein distance in RPN to model elliptical object. Moreover, Dong et al.~\cite{Dong} proposed Ellipse R-CNN, a two-stage detector with ellipse regression upon the Mask R-CNN~\cite{He_2017_ICCV} for detecting elliptical objects in occluded and cluttered scenes. However, all the above ellipse detection methods are anchor-based detection frameworks, which typically introduces many hyperparameters and usually difficult to train. Further, these methods highly rely on the quality of densely predefined anchors that need sufficient overlaps with ground truth. Very few works have attempted to employ anchor-free approaches~\cite{Zhou2019c,Yang2020a} to detect elliptical objects. 

\begin{figure}[t]
\includegraphics[width=\textwidth]{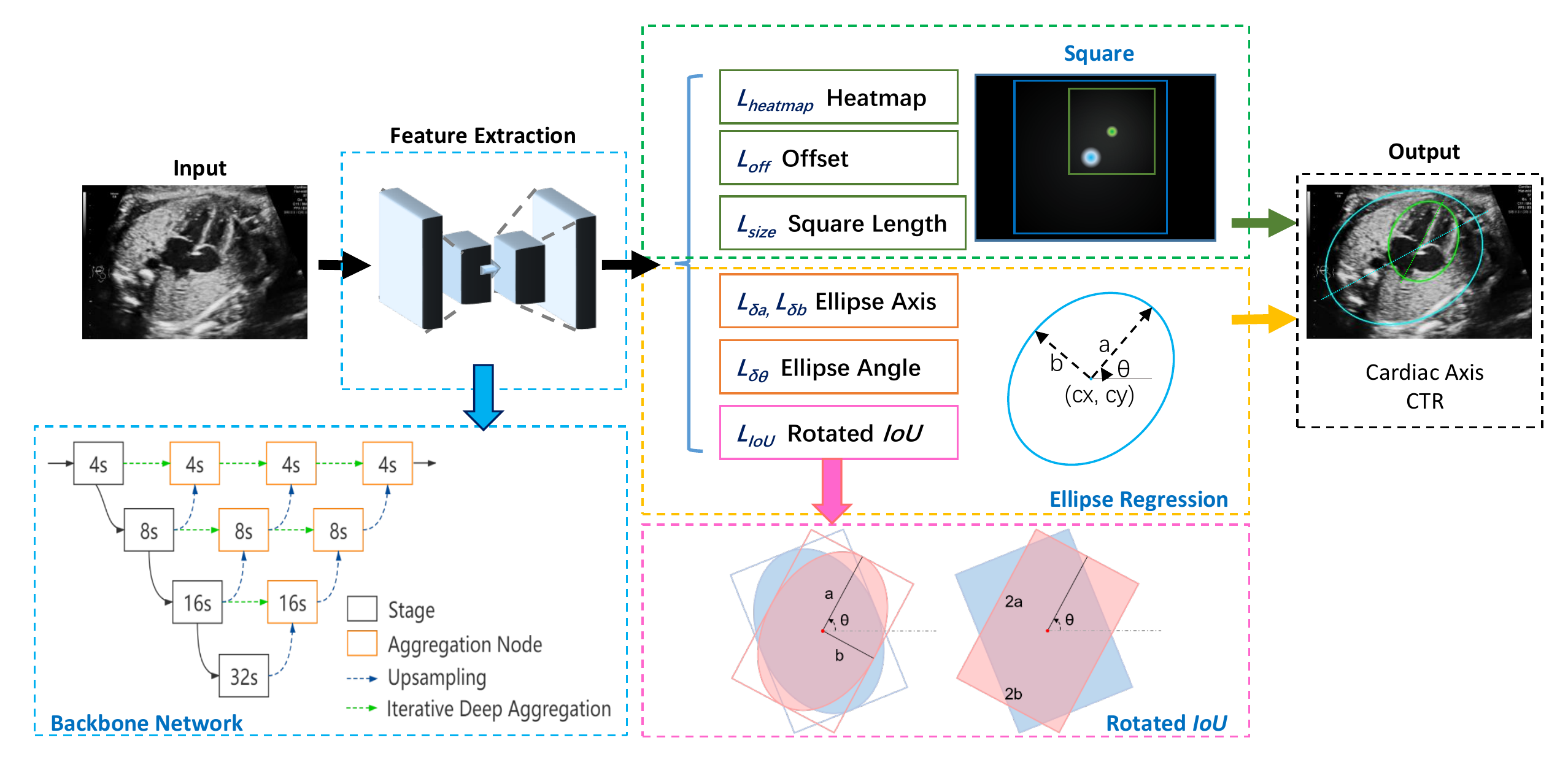}
\caption{Overview of the proposed ellipse detection network. EllipseNet consists of the feature extraction module, which is identical to deep layer aggregation network DLA-34~\cite{Yu_2018_CVPR}
; the square detection module, which detects the centers of the objects; and the ellipse regression module, where a rotated IoU loss is integrated.} \label{fig:network_archi}
\end{figure}

\section{Method}

\subsection{Overview of the framework}
The overview of our proposed EllipseNet is shown in Fig.~\ref{fig:network_archi}. It composes of a backbone network for feature extraction, an anchor-free square detection module and an ellipse regression module. All the above modules are simultaneously trained in an end-to-end manner. The backbone network and the square detection module are built upon CenterNet~\cite{Zhou2019c}, which is a keypoint-based anchor-free detection framework with high performance on center point localization.       
 

\subsection{Square Detection}
It is not stable and difficult to train if directly regressing the five parameters of ellipse. Inspired by Ellipse R-CNN~\cite{Dong}, we extend the elliptical region to a square with length $l=2\sqrt{a^2 + b^2}$, which shares the same center point with the ellipse. The extended square is closely related to the location and size of the ellipse but independent of the rotation angle $\theta$. Further, the degrees of freedom of the extended square is 1 (length $Q_l$ for shape) when its center point $(Q_x, Q_y)$ is learning, so that it is easier to optimize by the detection network. We therefore use the square detection module for ellipse localisation.

The square detection module consists of three heads: the center point heatmap head, the center point offset head and the square length head. Given an input image $I \in R^{W \times H \times 3}$ with height H and width W. The heatmap head predict the heatmap $\hat{Y} \in [0, 1]^{\frac{W}{R} \times \frac{H}{R} \times 2}$, representing the probability map of the center localization of thoracic and cardiac regions respectively. Here, we set output stride $R=4$ as suggested in~\cite{Zhou2019c}, where the square length head only predicts a square length but not the height and width as that within bounding box. We use pixel-wise regression with focal loss~\cite{Lin2017a} to optimize the center point heatmap and L1 loss to optimize the offset and square length. The heatmap loss is presented:

\begin{equation}
    L_{heatmap}=-\frac{1}{N} \sum_{x y c}\left\{
    \begin{array}{cc}
        \left(1-\hat{Y}_{x y c}\right)^{\alpha} \log \left(\hat{Y}_{x y c}\right) & \textrm { if } Y_{x y c}=1 \\ 
        \begin{array}{c}
            \left(1-Y_{x y c}\right)^{\beta}\left(\hat{Y}_{x y c}\right)^{\alpha} \log \left(1-\hat{Y}_{x y c}\right) \\
        \end{array} & \textrm { otherwise } \\
    \end{array}\right.   
\label{equation1}
\end{equation}
where $xyc$ represents the position in heatmap, $\alpha$ and $\beta$ are hyper-parameters of focal loss, where $\alpha = 2 $ and $\beta = 4$. The loss of square detection is formulated as:
\begin{equation}
    L_{Q} = L_{heatmap} + \lambda_{size} L_{size} + \lambda_{off} L_{off} .
\end{equation}

\subsection{Ellipse Regression}
When the square parameters $Q = (Q_x, Q_y, Q_l)$ are determined in the square detection module, the center points of the ellipse will be coarsely decided. Then the regression of the ellipses' parameters including semi-major axis $E_a$, semi-minor axis $E_b$, and rotation angle $E_{\theta}$ can be formulated as below:
\begin{equation}
    \begin{array}{c}
        \delta_a = (E_a / Q_l), \qquad 
        \delta_b = (E_b / Q_l), \qquad 
        \delta_\theta = \theta / \pi  \\
    \end{array} \label{eq:ellipse_param}
\end{equation}

\begin{equation}
    \theta = \left\{ \begin{array}{lc}
            \textrm{atan2}(\sin E_\theta, \cos E_\theta) & \qquad \textrm{if }\cos E_\theta \geq 0 \\
            \textrm{atan2}(- \sin E_\theta, -\cos E_\theta) & \qquad \textrm{if }\cos E_\theta < 0  \\
        \end{array}\right. 
\end{equation}
where $\delta_a$ and $\delta_b$ represent the ratios of $E_a$ and $E_b$ to the length of corresponding extended square, respectively. Here, we define the range of rotation angle $E_{\theta} \in (–\pi/2, \pi/2]$, and the difference between $-\pi/2$ and $\pi/2$ is 0 rather than $\pi$. We use separate heads to regress the parameters in Eq.~\ref{eq:ellipse_param} with L1 losses, and the ellipse regression loss $L_{E}$ is defined as a weighted sum of those losses:

\begin{equation} \label{eq:ellipse_loss}
    L_{E} = L_{\delta_a} + L_{\delta_b} + \lambda_{\delta_\theta} L_{\delta_\theta}
\end{equation}


\subsection{Rotated IoU Loss}

To further constrain the ellipse regression optimisation, we introduce an IoU metric, which is very popular in object detection tasks measuring the overlap between the boxes. More generally, the IoU loss can be formulated with an additional penalty term $L_{IoU} = 1 - IoU + P$, where $P$ further penalize the overlap between the predicted and the ground truth regions. For the original IoU loss, $P=0$. In this work, we chose the Distance-IoU loss~\cite{Zheng2019} as formulated below:
\begin{equation}
    L_{IoU} = 1 - IoU + P_{DIoU} = 1 - \frac{R_{pred} \cap R_{gt}}{R_{pred} \cup R_{gt}} + \frac{{\rho}^2}{{c}^2}.
\end{equation}
where a distance penalty term $P_{DIoU}$ is added to achieve faster convergence, $R_{pred}$ and $R_{gt}$ represent the predicted and groud truth regions, respectively, $\rho$ denotes the Euclidean distance between the predicted and ground truth center points, $c$ is the diagonal length of the smallest enclosing box covering the two  boxes. 

To compute the IoU loss of two rotated ellipses can be very expensive. Therefore, we introduce a rotated IoU~\cite{Zhou} to calculate the IoU of two rotated boxes tightly bounding the ellipses, as an approximate solution to the IoU of ellipses (see in Fig.~\ref{fig:network_archi}). It is worth mentioning that the bounding box shares the same center point and rotation angle with the corresponding ellipse, while the height and width are of the same length as major and minor axes of ellipse, respectively. 
To this end, all the parameters for ellipse regression are defined. The overall ellipse detection loss $L_{EN}$ can be formulated as:
\begin{equation} \label{eq:total_loss}
    L_{EN} = \lambda_Q L_Q + \lambda_E L_E + \lambda_{IoU} L_{IoU} .
\end{equation}
where $\lambda_{Q}$, $\lambda_{E}$, and $\lambda_{IoU}$ are the weighting parameters. 

\section{Experiments}

\subsection{Data acquisition}
The dataset consists of 2086 2D fetal echocardiographic images in 4-chamber view. The data were obtained from pregnant women aged from 19 to 45 at a gestational age of 18–39 weeks. The median age at pregnancy was 25±3 years, while the median gestational age was 29.4 weeks. The machine models include the GE Voluson E8 and E10 and the Philips EPIQ 7C. All the subjects used in this study are with ethical committee approval. The cardiac and thoracic regions are labelled in ellipses using an offline ellipse annotation tool provided by VGG Image Annotator\footnote{\url{https://www.robots.ox.ac.uk/~vgg/software/via/}}~\cite{dutta2019vgg} by two trained operators and examined by fetal echocardiographers. 

\subsection{Implementation Details}
We randomly select $80\%$ of the images for training and the remainder for testing. All the images are resized to $896 \times 608$ before feeding into the network. Random left-right flipping, scaling, shifting and random Gaussian noise are added for data augmentation.
We use deep layer aggregation network~\cite{Yu_2018_CVPR} DLA-34 with output stride 4 as the backbone network, and the output heads are defined as shown in Fig.~\ref{fig:network_archi}. Before each output head, we use a $3 \times 3$ convolutional layer with 256 channels followed by a ReLU layer. Moreover, we add a sigmoid activation function to the end of ellipse axis heads for predicting the ratios $\delta_a$ and $\delta_b$. The network is trained by minimizing the loss function in Eq.~\ref{eq:total_loss} with ADAM optimizer. All loss weights $\lambda_*$ are set to 1 except $\lambda_{size}=0.1$ and $\lambda_{\delta_\theta} = 5$. We train our model with batch size 24 and initial learning rate 5e-4 for 120 epochs. All experiments were performed on a single NVIDIA Tesla V100 GPU in PyTorch. 

\subsection{Results}
To compute the overlap between rotated ellipses, we draw the ellipses as shape masks and calculating the dice coefficient between masks to evaluate the detection performance. The CTR is formulated as ${R} = {b_C} / {b_T}$, where $b_C$ represents the length of minor axis of the cardiac ellipse, $b_T$ represents that of the thoracic ellipse. Then, the CTR precision can be defined as: ${P_{avg}} = 1 - \frac{ |R_{true} - R_{pred} |}{R_{true}}$, where $R_{pred}$ and $R_{true}$ denote the  predicted CTR and the ground truth CTR, respectively.

\renewcommand\tabcolsep{5.0pt}
\begin{table}
\caption{Experimental results between different configurations and comparison results. The dice coefficient of cardiac and thoracic regions, the average of them and the precision of estimated CTR are reported here.}
\centering
\begin{tabular}{cccccc}
\hline
Methods &  Setting & $Dice_T$ & $Dice_C$ & $Dice_{all}$ & $P_{avg}$\\
\hline
\hline
Segment+ellipse-fit \cite{Sinclair2018} & Residual U-Net & 0.8750 & 0.9182 & 0.9112 & 0.8520\\
CircleNet \cite{Yang2020a} & Hourglass & 0.8966 & 0.8666 & 0.8816 & 0.8555\\
CircleNet \cite{Yang2020a} & DLA & 0.8966 & 0.8729 & 0.8847 & 0.8614\\
\hline
EllipseNet (Ours) & only IoU loss & 0.8813 & 0.8520 & 0.8666 & 0.8855 \\
EllipseNet (Ours) & w/o IoU loss & 0.9338 & 0.9108 & 0.9224 & 0.8841 \\
EllipseNet (Ours) & w/ IoU loss & \textBF{0.9430} & \textBF{0.9224} & \textBF{0.9336} & \textBF{0.8949}\\
\hline
\end{tabular} \label{table1}
\end{table}

\begin{figure*}[htb]
  \centering
  \subfigure[Ground Truth]{
  \begin{minipage}[b]{0.23\linewidth}
    \includegraphics[width=1\linewidth]{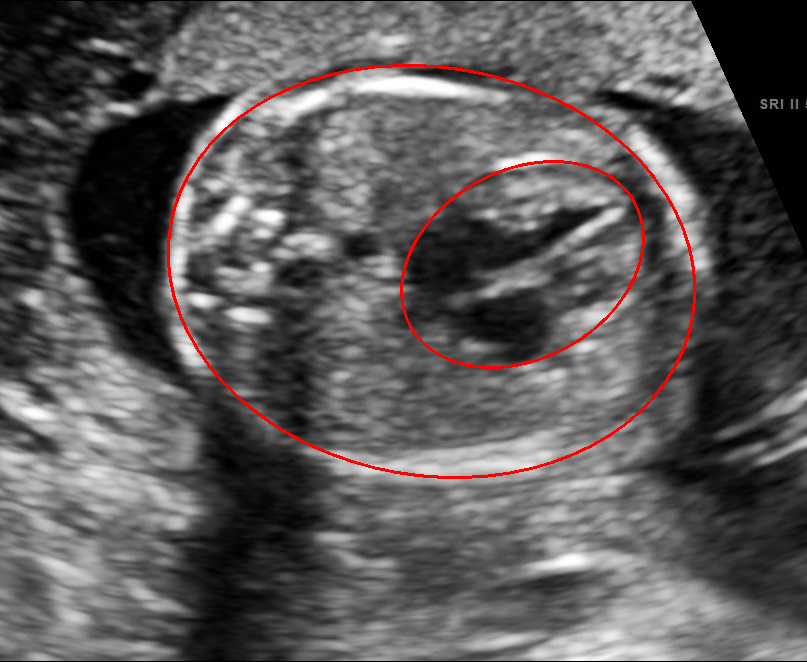} \\
    \includegraphics[width=1\linewidth]{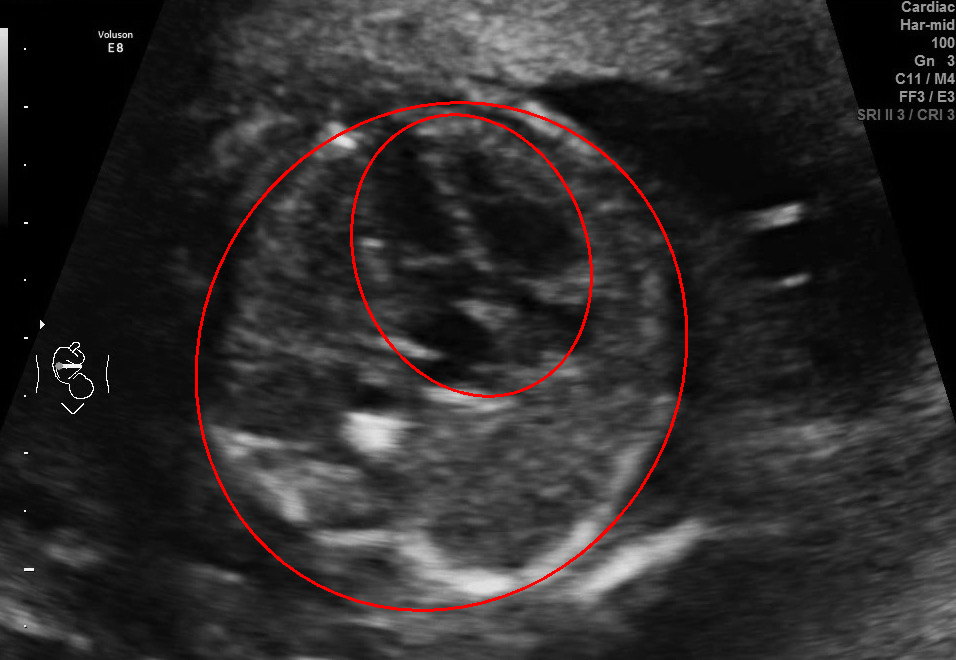} \\ 
    \includegraphics[width=1\linewidth]{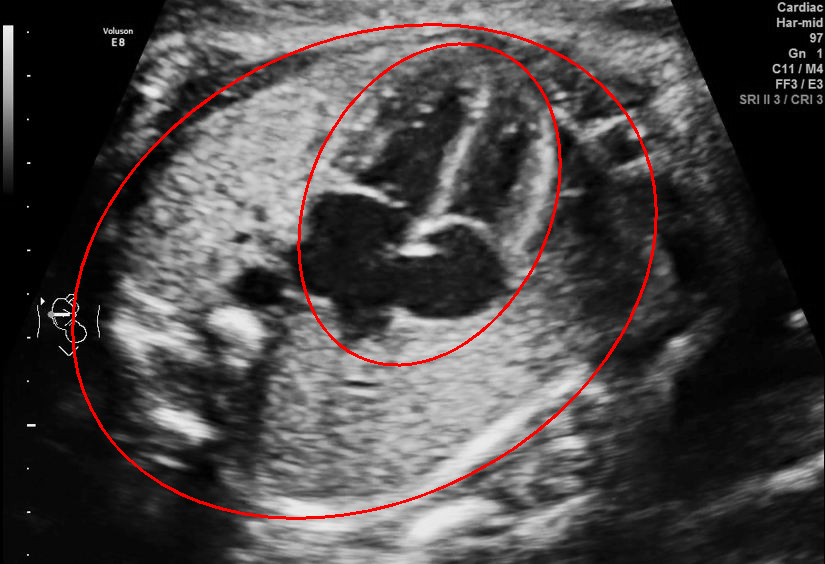} 
  \end{minipage}}
  \subfigure[Ellipse-fit]{
  \begin{minipage}[b]{0.23\linewidth}
    \includegraphics[width=1\linewidth]{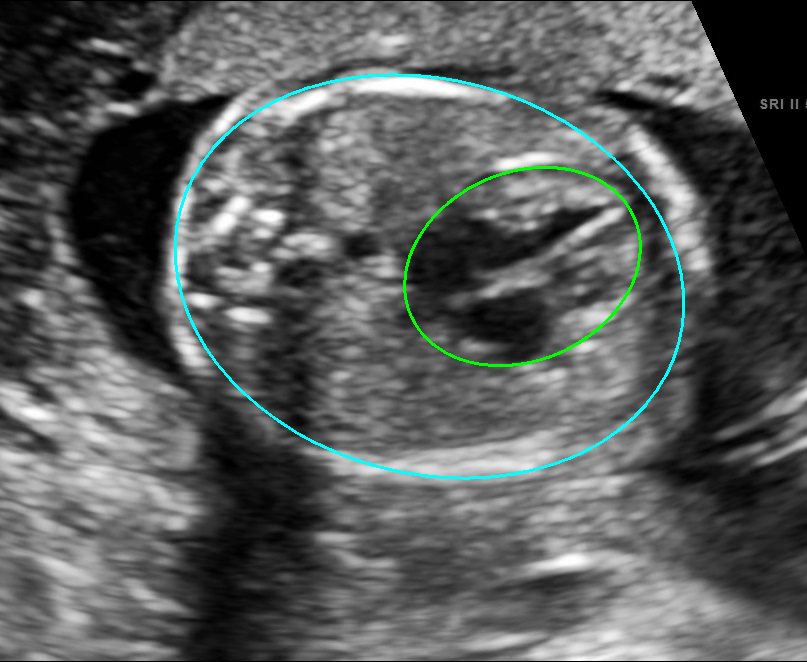} \\ 
    \includegraphics[width=1\linewidth]{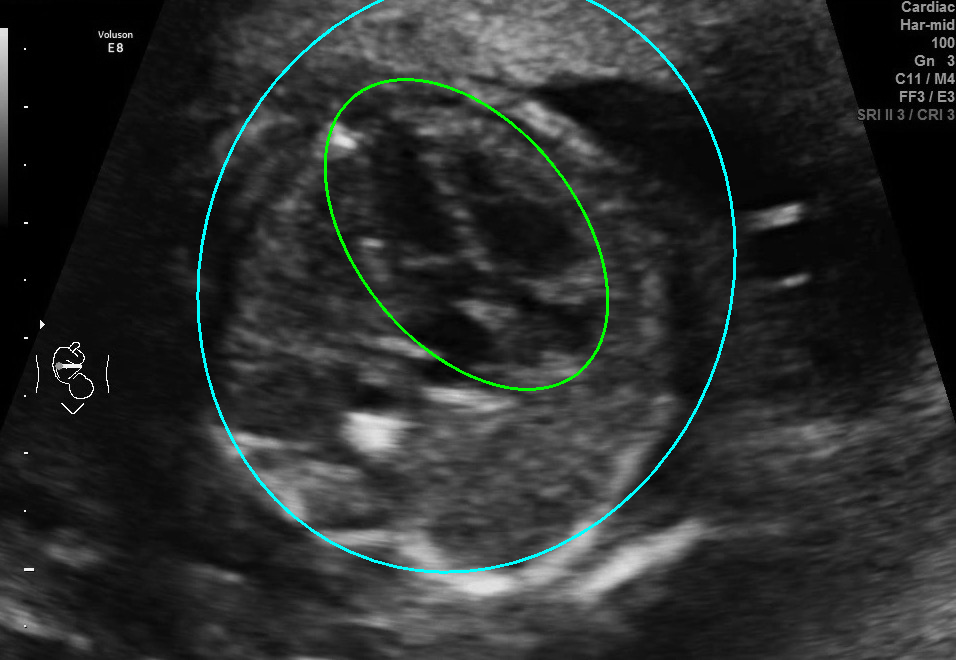} \\ 
    \includegraphics[width=1\linewidth]{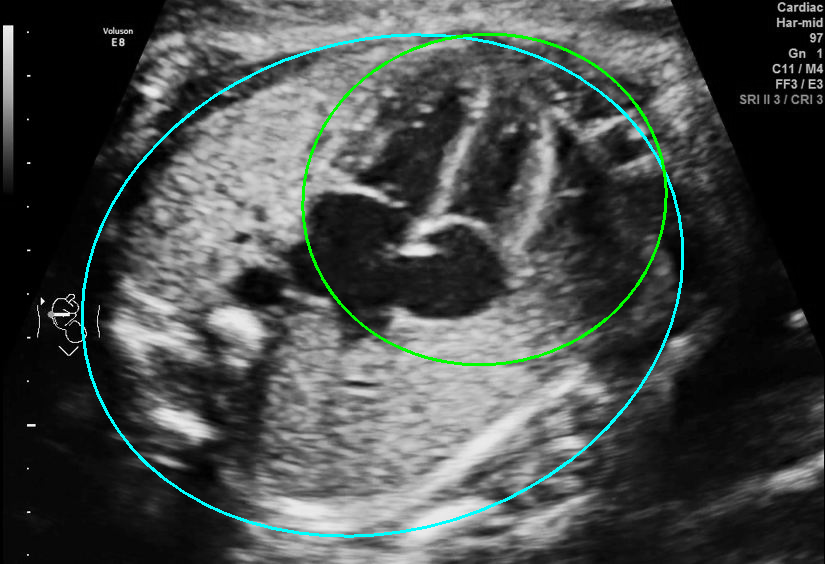}
  \end{minipage}}
  \subfigure[CircleNet]{
  \begin{minipage}[b]{0.23\linewidth}
    \includegraphics[width=1\linewidth]{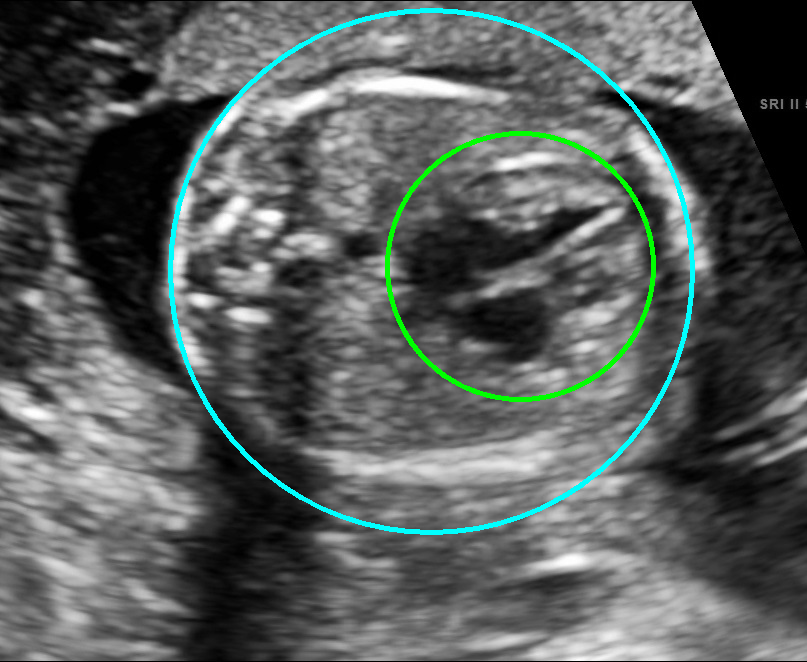} \\ 
    \includegraphics[width=1\linewidth]{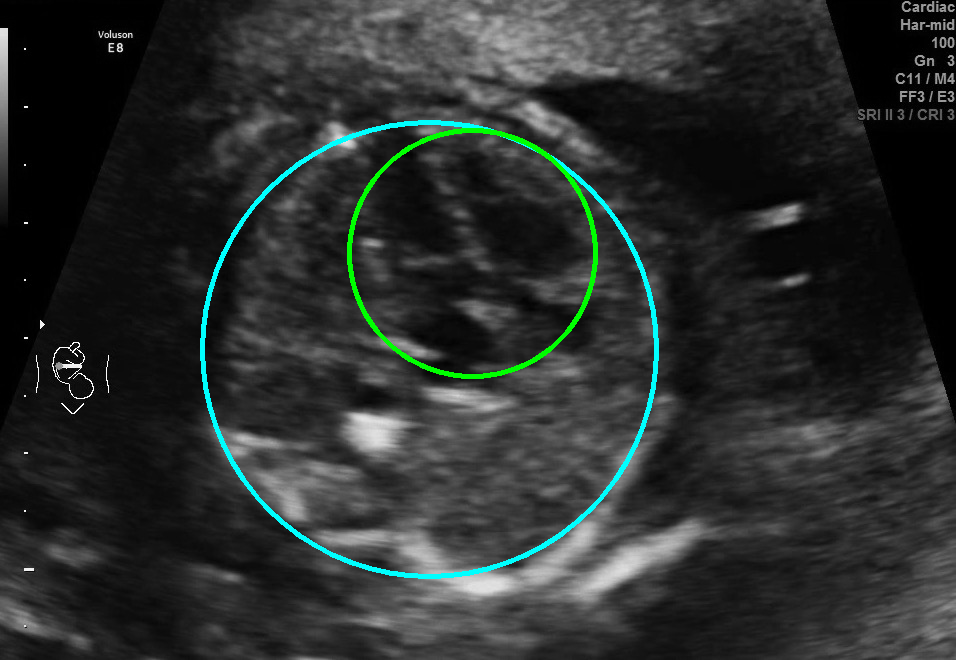} \\ 
    \includegraphics[width=1\linewidth]{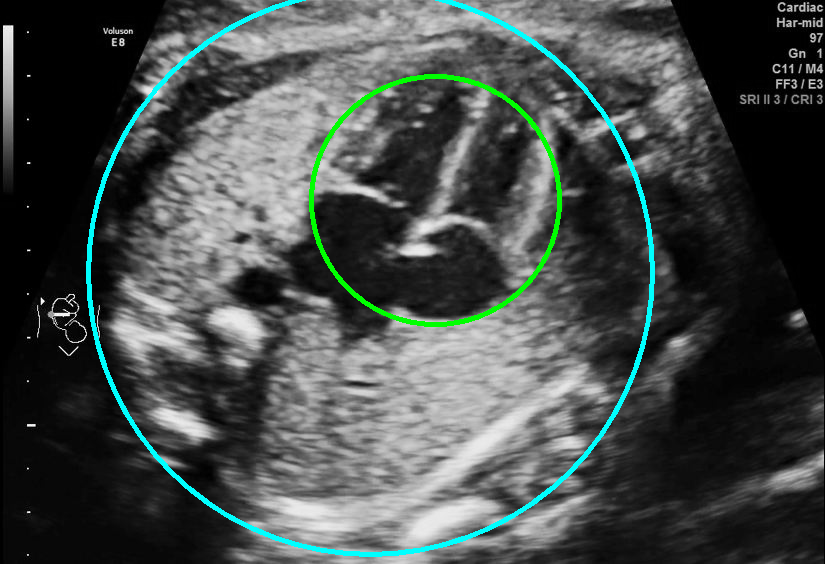}
  \end{minipage}}
  \subfigure[Ours]{
  \begin{minipage}[b]{0.23\linewidth}
    \includegraphics[width=1\linewidth]{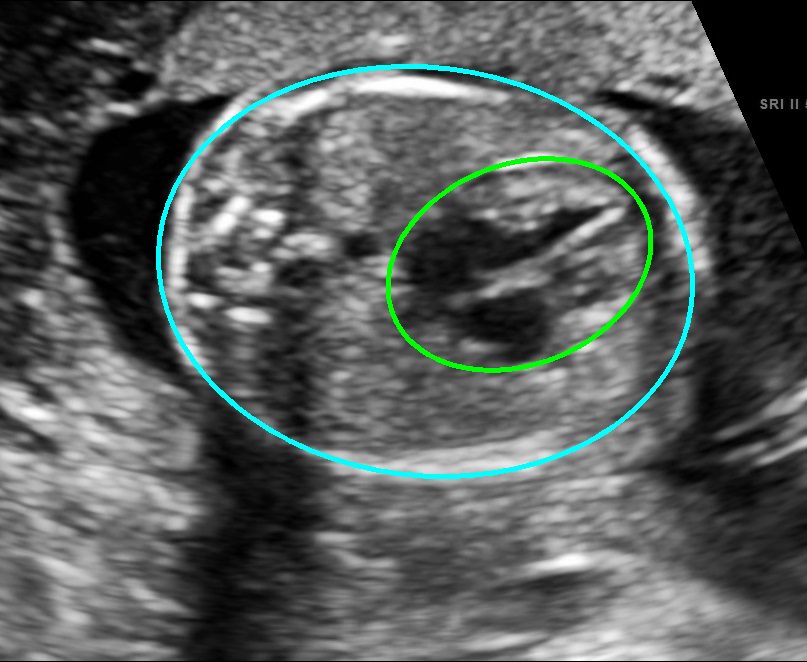} \\ 
    \includegraphics[width=1\linewidth]{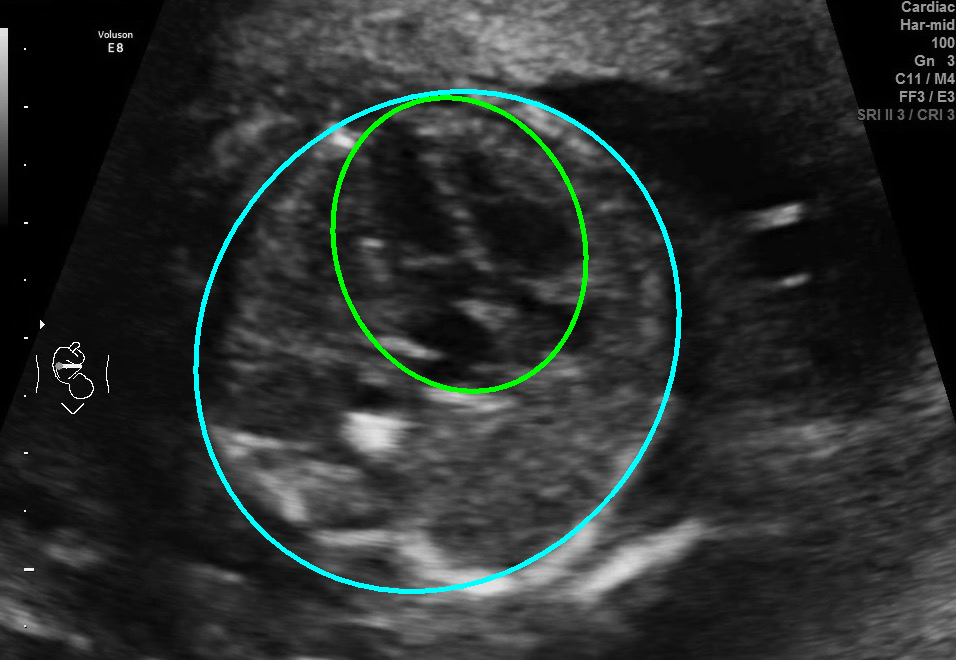} \\ 
    \includegraphics[width=1\linewidth]{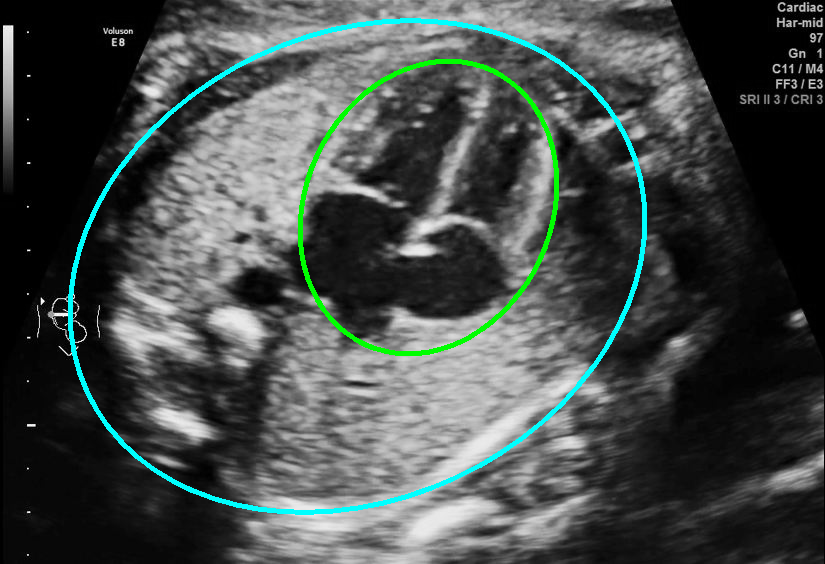}
  \end{minipage}}
\caption{\textBF{Visualization results of different methods.} The columns from left to right display the ground truth and prediction results from segment-based ellipse-fit method, CircleNet and our proposed EllipseNet with IoU loss. The red ellipses represent ground truth of cardiac and thoracic regions, while green and cyan ellipses represent the predicted cardiac and thoracic region respectively. }
	\label{fig:visualize}
\end{figure*}
\begin{figure*}[htb] 
  \centering
  \subfigure[only IoU loss]{
  \begin{minipage}[b]{0.25\linewidth}
    \includegraphics[width=1\linewidth]{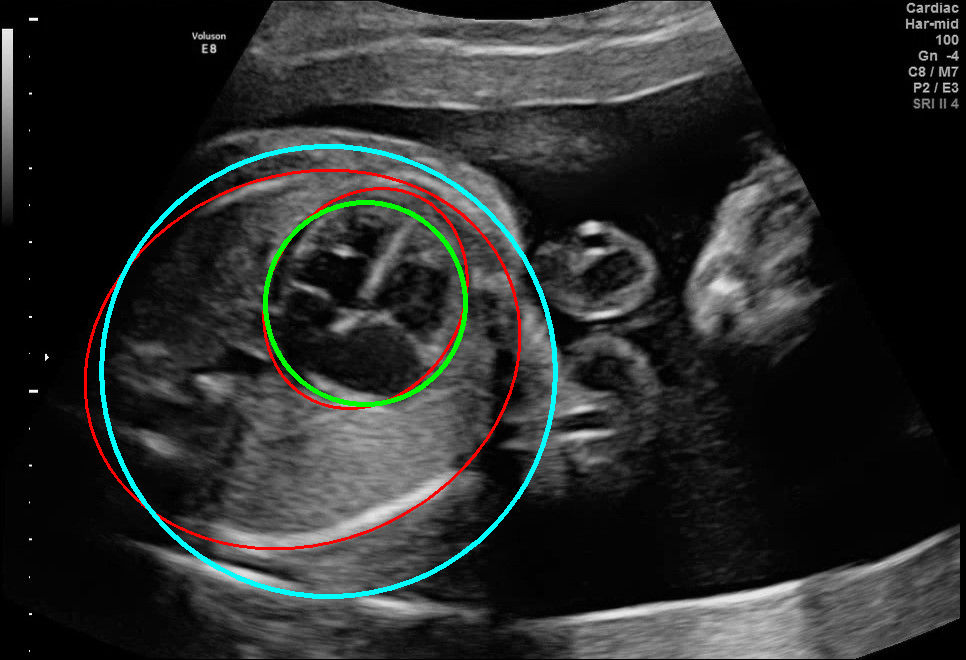}  \\
    \includegraphics[width=1\linewidth]{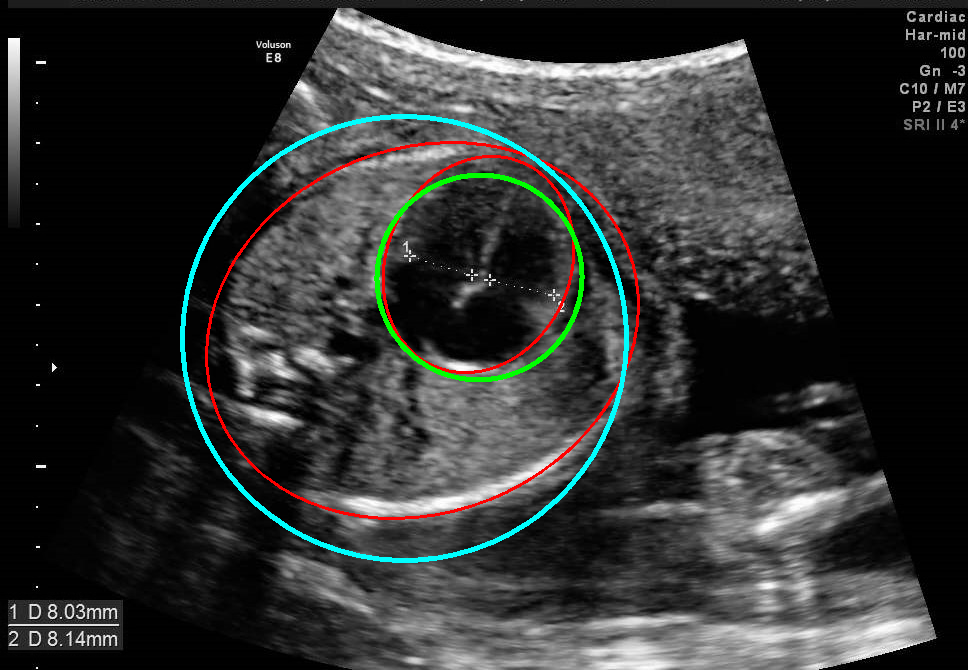}  \\ 
    \includegraphics[width=1\linewidth]{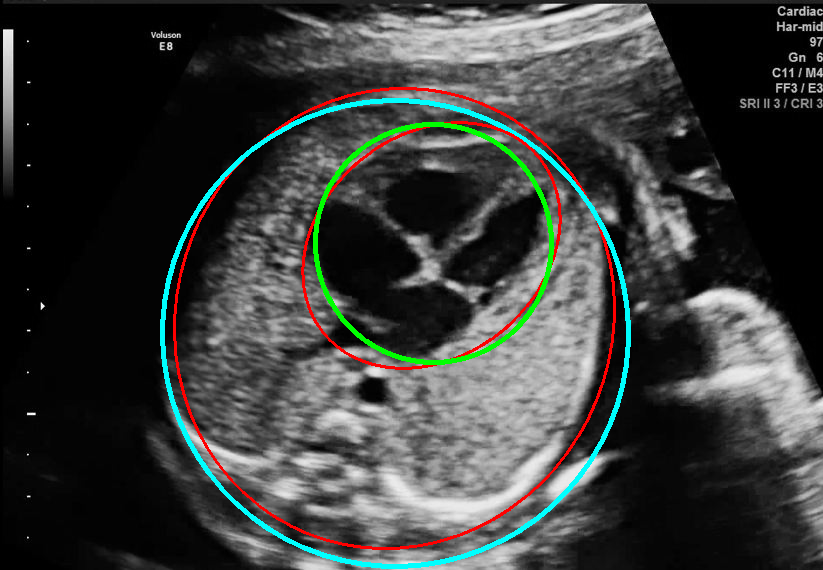} 
  \end{minipage}}
  \subfigure[w/o IoU loss]{
  \begin{minipage}[b]{0.25\linewidth}
    \includegraphics[width=1\linewidth]{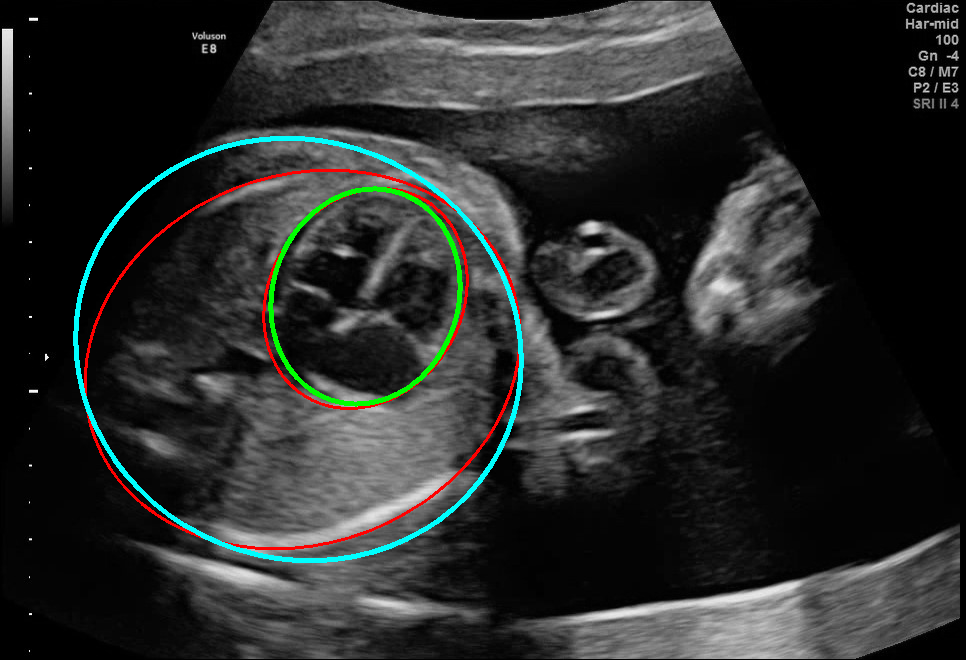} \\ 
    \includegraphics[width=1\linewidth]{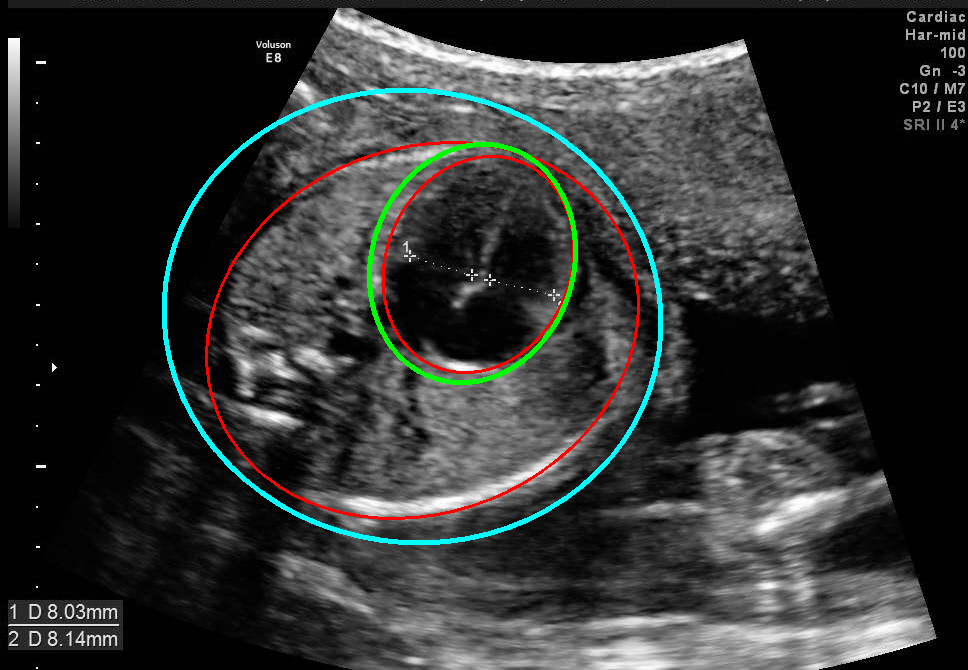} \\ 
    \includegraphics[width=1\linewidth]{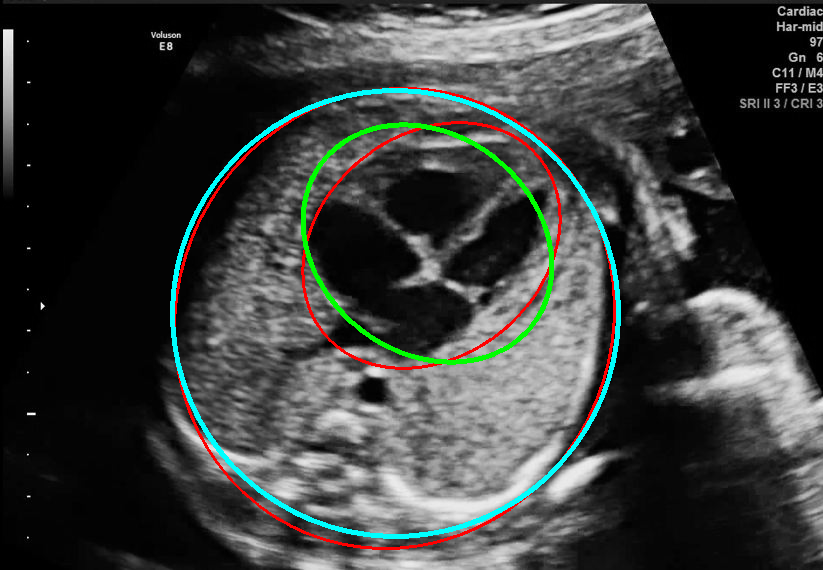}
  \end{minipage}}
  \subfigure[w/ IoU loss]{
  \begin{minipage}[b]{0.25\linewidth}
    \includegraphics[width=1\linewidth]{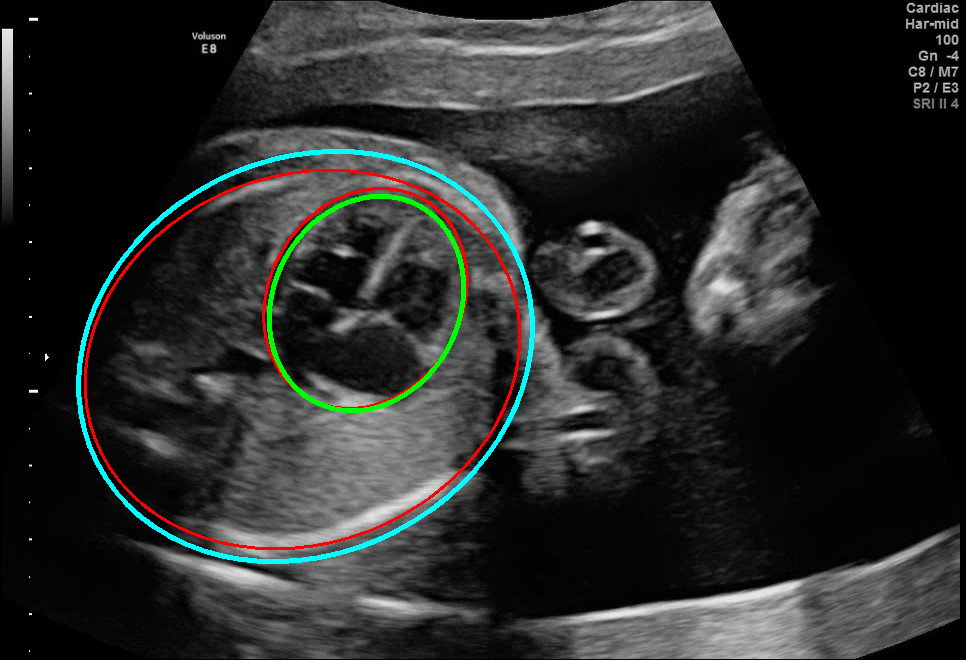}  \\ 
    \includegraphics[width=1\linewidth]{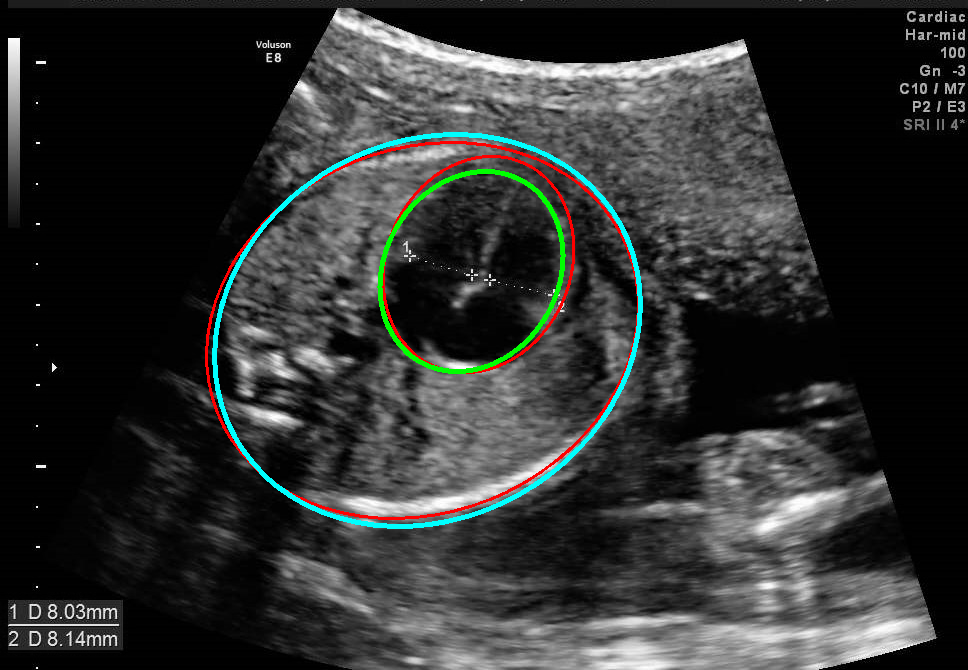}  \\
    \includegraphics[width=1\linewidth]{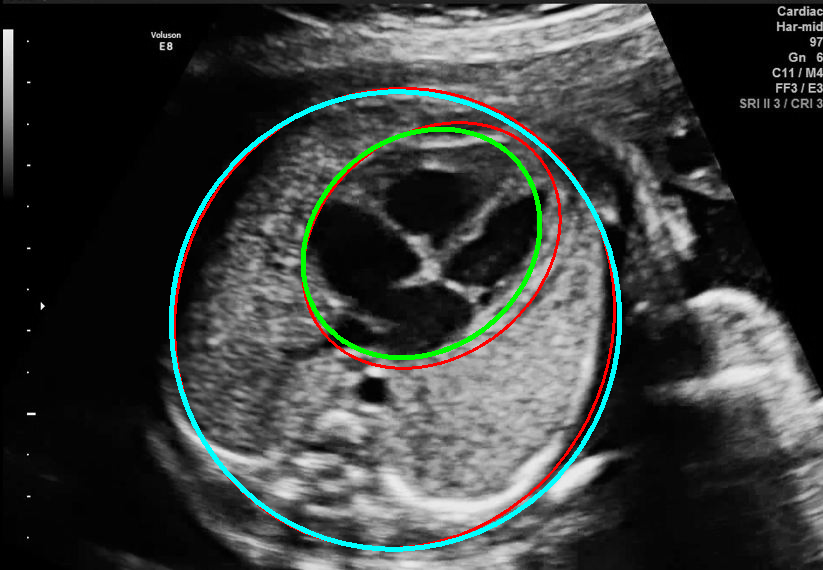} 
  \end{minipage}}
  \caption{\textBF{Results of ablation study.} The columns display results from different configurations of our proposed method. For better visualization, predictions and ground truth are drawn in the same image.}
	\label{fig:ablation}
\end{figure*}

\subsubsection{Comparison with other methods}
We compare EllipseNet with several automatic estimation-related detection methods to demonstrate its effectiveness. First, we compare it with~\cite{Sinclair2018}, which trains a Fully Convolutional Network (FCN) segmentation model to obtain the structure contour and then uses an ellipse to fit it. Instead of using the FCN in the original paper, we use MONAI\footnote{\url{https://monai.io/}} to train a more powerful U-Net~\cite{Kerfoot2019} as the segmentation network for better performance. Second, we compare it with CircleNet~\cite{Yang2020a}, an anchor-free detection method with circle representation. The circle can be regarded as a special case of an ellipse with the same radius for all points, so the CTR can be calculated by the radii of circles. It should be noted that CircleNet cannot be used for cardiac axis measurements since it treats the objects as circles. As presented in Table.~\ref{table1}, our proposed method outperforms the others with all the dice coefficients and CTR precision. 

Although the segmentation model trained on the fetal ultrasound dataset achieves a relative high average dice scores averaging over 0.9, the fitted ellipse highly depends on the segmentation results. As shown in the middle row of Fig.~\ref{fig:visualize}, the segment-based ellipse-fit method performs well when the image quality is good (first row), but the performance degrades when the segmentation is affected by image artifacts such as the acoustic shadowing (second and third row). Our proposed method is more robust to image quality and shadows. We also tried to compare to the GPN~\cite{Li2019c} with their open source code on our dataset, however, the results are not comparable to ours and those presented in Table.~\ref{table1}. It is difficult to conclude whether it is caused by the network itself or the training strategies. We, therefore, did not include the comparative results in this work.
\subsubsection{Ablation Study.}
In this section, we evaluate the effect of different loss terms in EllipseNet by ablation study. As shown in the lower part of Table.~\ref{table1}, IoU loss improves detection performance remarkably. EllipseNet with IoU loss achieves a gain of 0.92 and 1.16 points in cardiac and thoracic dice coefficients respectively. On the other hand, EllipseNet with only IoU loss (without ellipse regression loss defined in Eq.~\ref{eq:ellipse_loss}) does not achieve satisfactory performance, which demonstrate the necessarity to include the paremeters' regression losses. It worth mentioning that the exact IoU of two rotated ellipses is always greater than that of corresponding rotated bounding boxes based on our observation. The IoU loss brings promotion in terms of bounding box IoU, which can be regarded as a lower bound of the ellipse IoU. This may explain why IoU loss is effective for ellipse regression in another view. 

As shown in Fig.~\ref{fig:ablation} , both the proposed ellipse regression loss and IoU loss are necessary for ellipse detection. If the EllipseNet only supervised by IoU loss (first column), the model fails to optimize the major and minor axis separately, and the predicted ellipses degenerate into circles like the CircleNet. It is clear that the supervision of IoU loss can help to improve the prediction of location and shape (first and second row) and to correct the angle (last row).
\section{Conclusion}
In this work, we presented an anchor-free ellipse detection network, named EllipseNet, for automatic measurement of fetal cardiac biometrics in 4-chamber view in echocardiographic screening. The proposed framework is anchor-free one-stage network that can perform ellipse detection and CTR calculation automatically with remarkably high precision. The pipeline can be easily adapted to other elliptical object detection and biometric measurement tasks.

\section*{Acknowledgement}
This work is supported by Ministry of Science and Technology of China - Peng Cheng Laboratory Special Project (grant No. PCNL2021ZDXM06), and in part by the Beijing Municipal Science and Technology Commission under Grant Z181100001918008.

\bibliographystyle{splncs}
\bibliography{reference.bib} 

\end{document}